\documentstyle[pra,twocolumn,aps]{revtex}

\begin{document}
\draft
\title{Phase preparation by atom counting of Bose-Einstein condensates in mixed
states.}
\author{R. Graham\thanks{%
Permanent address: Universit\"{a}t GH Essen, Fachbereich Physik, D45117
Essen, Germany.}, T. Wong, M. J. Collett, S. M. Tan and D. F. Walls}
\address{Department of Physics, University of Auckland, Private Bag 92019,\\
Auckland,\\
New Zealand}
\maketitle

\begin{abstract}
We study the build up of quantum coherence between two Bose-Einstein
condensates which are initially in mixed states. We consider in detail the
two cases where each condensate is initially in a thermal or a Poisson
distribution of atom number. Although initially there is no relative phase
between the condensates, a sequence of spatial atom detections produces an
interference pattern with arbitrary but fixed relative phase. The visibility
of this interference pattern is close to one for the Poisson distribution of
two condensates with equal counting rates but it becomes a stochastic
variable in the thermal case, where the visibility will vary from run to run
around an average visibility of $\pi /4.$ In both cases, the variance of the
phase distribution is inversely proportional to the number of atom
detections in the regime where this number is large compared to one but
small compared with the total number of atoms in the condensates.
\end{abstract}


\section{Introduction}

The recent experimental realization of weakly interacting Bose-Einstein
condensates \cite{anderson,bradley,davis} has stimulated a large amount of
theoretical work on the properties of these condensates. Recently, there has
been a great interest in the interference and the establishment of a
relative phase\cite{Jan,Nar,Cas,Cir,Bar,Jac,Wong,Mol} between Bose-Einstein
condensates which start in Fock states or as a mixture of coherent states.
In this paper we show by numerical and analytical calculations how a
relative phase is established between two independent condensates initially
in mixed states. In particular we look at the quantum interference when the
condensates begin in thermal or Poisson mixed states. There is a significant
difference in the visibility of the interference patterns between these two
cases.

\section{Quantum Phase between initial Mixtures}

We consider two Bose-Einstein condensates which are dropped on top of each
other. This system was first proposed by {Javanainen and Yoo\cite{Jan}. }An
example of such a system is a hot wire grid placed below the condensates
which acts as an atom detector since it removes atoms from the condensates
as they fall under the influence of gravity. {Javanainen and Yoo showed that
a spatial interference pattern would be observed illustrating the presence
of a relative phase between the falling }condensates. We will show here that
this relative phase is also present when the initial states of the
condensates are no longer a pure Fock state of known number but are a
mixture of number states. This corresponds physically to the situation where
we are uncertain of the initial number of atoms in each condensate. We shall
take this mixture to have a number distribution $P_{n}$ giving an initial
density operator of the form 
\begin{equation}
\rho =\sum_{n_{1}=0}^{\infty }\sum_{n_{2}=0}^{\infty
}P_{n_{1}}|n_{1}\left\rangle {}\right\langle n_{1}|\otimes
P_{n_{2}}|n_{2}\left\rangle {}\right\langle n_{2}|,  \label{densityopr}
\end{equation}
where $n_{1}$ and $n_{2}$ refer to the first and second condensate
respectively. The spatial interference is established since when we detect
the atoms from the condensates we are unaware which condensate the atom came
from. This interference pattern is built up from the spatial detections of
individual atoms. After $m$ detections we observe atoms at positions $%
\left\{ x_{1},\ldots ,x_{m}\right\} $. If we denote the field operator for
the detection of an atom at $x$ as $\hat{\psi}\left( x\right) $ then the
joint probability of $m$ detections at the $\left\{ x_{1},\ldots
,x_{m}\right\} $ positions is 
\begin{eqnarray}
p^{\left( m\right) }(x_{1},\ldots ,x_{m}) &=&{\cal N}^{\left( m\right) }{\bf %
Tr}\left\{ \rho \hat{\psi}^{\dag }(x_{1})\ldots \hat{\psi}^{\dag
}(x_{m})\right.  \nonumber \\
&&\left. \times \hat{\psi}(x_{m})\ldots \hat{\psi}\left( x_{1}\right)
\right\} ,  \label{TraceEq}
\end{eqnarray}
where the symbol {\bf Tr} denotes the trace over the $n_{1}$ and $n_{2}$
number states. The normalization ${\cal N}^{\left( m\right) }$ is defined by 
\begin{eqnarray}
{\cal N}^{\left( m\right) } &=&\left[ \int dx_{1}\ldots dx_{m}{\bf Tr}%
\left\{ \rho \hat{\psi}^{\dag }(x_{1})\ldots \hat{\psi}^{\dag
}(x_{m})\right. \right.  \nonumber \\
&&\left. \times \left. \hat{\psi}(x_{m})\ldots \hat{\psi}\left( x_{1}\right)
\right\} \right] ^{-1}
\end{eqnarray}
which is independent of $x_{1},\ldots ,x_{m}$ but will in general depend on $%
\rho .$ We define the field operator for the two condensate system\cite
{Jan,Wong} as 
\begin{equation}
\hat{\psi}(x)=\hat{a}_{1}+\sqrt{\Gamma }\hat{a}_{2}e^{-i\phi (x)},
\label{fieldopr}
\end{equation}
with 
\begin{equation}
\phi (x)=(k_{1}-k_{2})x
\end{equation}
where $\hat{a}_{1}$ and $\hat{a}_{2}$ are the atom annihilation operators
for the first and second condensate respectively with each condensate
possessing momentum $k_{1}$ and $k_{2}$. We define the ratio $\Gamma =\gamma
_{2}/\gamma _{1}$ where $\gamma _{1}$ and $\gamma_{2}$ are the detection
rates for each of the condensates. We shall assume $\gamma_{1}\geq
\gamma_{2}$ without restriction of generality.

Substituting equation (\ref{densityopr}) into~(\ref{TraceEq}), we obtain the
following expression for the joint probability. 
\begin{eqnarray}
&&p^{\left( m\right) }(x_{1},\ldots ,x_{m})  \nonumber \\
&=&{\cal N}^{\left( m\right) }\sum_{n_{1}=0}^{\infty }\sum_{n_{2}=0}^{\infty
}P_{n_{1},n_{2}}\langle n_{1},n_{2}|\hat{\psi}^{\dag }(x_{1})\ldots \hat{\psi%
}^{\dag }(x_{m})  \nonumber \\
&&\times \hat{\psi}(x_{m})\ldots \hat{\psi}(x_{1})|n_{1},n_{2}\rangle
\label{jointprob}
\end{eqnarray}
where we have written $P_{n_{1},n_{2}}=P_{n_{1}}P_{n_{2}}$ for convenience.
The expression for the joint probability above shows that it is a weighted
sum of probabilities over fixed numbers $n_{1}$ and $n_{2}$ of initial
numbers of atoms in each condensate. The weighting is determined by the
probability distribution of the number of atoms. For very narrow
distributions, corresponding to an initial fixed number of atoms in each
condensate, we see that this will simplify down to the expression used
previously by Javanainen and Yoo \cite{Jan}. However for broader
distributions, for example a thermal distribution, this sum will affect this
probability and hence the spatial interference.

\subsection{Visibility conditioned on $1$ detection}

Before we show how to numerically generate this interference pattern, we
look at the build up of interference for the first two detections. Let us
start by considering a joint Fock state. After one detection in which an
atom is observed at $x_{1}$, the un-normalized state vector for the system
is 
\begin{eqnarray}
\hat{\psi}(x_{1})|n_{1},n_{2}\rangle  &=&\sqrt{n_{1}}|n_{1}-1,n_{2}\rangle  
\nonumber \\
&&+\sqrt{\Gamma n_{2}}e^{-i\phi (x_{1})}|n_{1},n_{2}-1\rangle .  \label{p1}
\end{eqnarray}
The joint probability density for detecting atoms at $x$ and $x_{1}$
starting from $|n_{1},n_{2}\rangle $ is then proportional to 
\begin{eqnarray}
\langle \varphi _{2}\left( \phi \right) |\varphi _{2}\left( \phi \right)
\rangle  &=&\langle n_{1},n_{2}|\hat{\psi}^{\dag }(x_{1})\hat{\psi}^{\dag
}(x)\hat{\psi}(x)\hat{\psi}(x_{1})|n_{1},n_{2}\rangle   \nonumber \\
&=&n_{1}(n_{1}-1)+\Gamma ^{2}n_{2}(n_{2}-1)  \nonumber \\
&&+2\Gamma n_{1}n_{2}\left\{ 1+\cos \left[ \phi (x)-\phi (x_{1})\right]
\right\} .
\end{eqnarray}
Since the initial state of interest is a mixture of Fock states with weights 
$P_{n_{1},n_{2}}$, the joint probability for this initial state by Eq. (\ref
{jointprob}) is 
\begin{eqnarray}
p^{\left( 2\right) }\left( x,x_{1}\right)  &=&{\cal N}^{\left( 2\right)
}\sum_{n_{1},n_{2}}P_{n_{1},n_{2}}\langle n_{1},n_{2}|\hat{\psi}^{\dag
}(x_{1})\hat{\psi}^{\dag }(x)  \nonumber \\
&&\times \hat{\psi}(x)\hat{\psi}(x_{1})|n_{1},n_{2}\rangle   \nonumber \\
&=&{\cal N}^{\left( 2\right) }\sum_{n_{1},n_{2}}P_{n_{1},n_{2}}\left(
n_{1}(n_{1}-1)+\Gamma ^{2}n_{2}(n_{2}-1)\right.   \nonumber \\
&&\left. +2\Gamma n_{1}n_{2}\left\{ 1+\cos \left[ \phi (x)-\phi
(x_{1})\right] \right\} \right)   \nonumber \\
&=&{\cal N}^{\left( 2\right) }\left( \left[ \langle n_{1}^{2}\rangle
-\langle n_{1}\rangle \right] +\Gamma ^{2}\left[ \langle n_{2}^{2}\rangle
-\langle n_{2}\rangle \right] \right.   \nonumber \\
&&\left. +2\Gamma \langle n_{1}\rangle \langle n_{2}\rangle \left\{ 1+\cos
\left[ \phi (x)-\phi \left( x_{1}\right) \right] \right\} \right) 
\end{eqnarray}
where we have used angle brackets to denote averages taken over $%
P_{n_{1},n_{2}}$. These factorize as we assume that $%
P_{n_{1},n_{2}}=P_{n_{1}}P_{n_{2}}.$ The conditional probability $%
p(x|x_{1})=p^{(2)}(x,x_{1})/p^{(1)}(x_{1})$ differs from the above only by a 
$x$-independent factor and we may write 
\begin{equation}
p(x|x_{1})=\frac{{\cal N}^{\left( 2\right) }}{{\cal N}^{\left( 1\right) }}%
\left\{ 1+{\cal V}\cos \left[ \phi (x)-\phi \left( x_{1}\right) \right]
\right\}   \label{eqn_vis}
\end{equation}
where 
\begin{equation}
{\cal V=}\frac{2\Gamma \langle n_{1}\rangle \langle n_{2}\rangle }{\left[
\langle n_{1}^{2}\rangle -\langle n_{1}\rangle \right] +\Gamma ^{2}\left[
\langle n_{2}^{2}\rangle -\langle n_{2}\rangle \right] +2\Gamma \langle
n_{1}\rangle \langle n_{2}\rangle }  \label{visibility}
\end{equation}
may be interpreted as the conditional visibility of the interference pattern.

For a thermal distribution we use the following relationship between the
second moment and the mean $\overline{m}\equiv \langle m\rangle ,$%
\begin{equation}
\langle m^{2}\rangle =2\overline{m}^{2}+\overline{m}.  \label{pro_eqn}
\end{equation}
Substituting the above equations into Eq.~(\ref{visibility}) we obtain an
expression for the conditional visibility in terms of the means 
\begin{equation}
{\cal V}_{{\rm thermal}}\left( \overline{n}_{1},\overline{n}_{2}\right) 
{\cal =}\frac{\Gamma \overline{n}_{1}\,\overline{n}_{2}}{\overline{n}%
_{1}^{2}+\left( \Gamma \overline{n}_{2}\right) ^{2}+\Gamma \overline{n}_{1}\,%
\overline{n}_{2}}
\end{equation}
which gives a maximum visibility for equal net detection rate in each
condensate, i.e. giving a maximum value of one third for $\overline{n}%
_{1}=\Gamma \overline{n}_{2}.$

Alternatively, for the case of a Poissonian number distribution we use the
relationship 
\begin{equation}
\langle m^{2}\rangle =\overline{m}^{2}+\overline{m}.  \label{property_eqn}
\end{equation}
Proceeding in the same manner as in the thermal case we obtain 
\begin{equation}
{\cal V}_{{\rm Poisson}}(\overline{n}_{1},\overline{n}_{2})=\frac{2\Gamma 
\overline{n}_{1}\,\overline{n}_{2}}{(\overline{n}_{1}+\Gamma \overline{n}%
_{2})^{2}},  \label{vis_poisson}
\end{equation}
which has a maximum value of one half for $\overline{n}_{1}=\Gamma \overline{%
n}_{2}.$ The values of one third and one half that we have obtained have
also been seen in optical experiments where intensity correlations were
measured for Poissonian and thermal light sources by Rarity {\em et. al.} 
\cite{Rarity}

Let us also consider the limiting case of fixed initial number, where we
know that there are exactly $n_{1}$ atoms in one condensate and $n_{2}$ in
the other. Then using Eq. (\ref{visibility}) the conditional visibility is 
\begin{equation}
{\cal V}_{{\rm Fock}}(n_{1},n_{2})=\frac{2\Gamma n_{1}\,n_{2}}{(n_{1}+\Gamma
n_{2})^{2}-\left( n_{1}+\Gamma ^{2}n_{2}\right) }.
\end{equation}
The maximum occurs again when $n_{1}=\Gamma n_{2}$ where the net detection
rates for both condensates are equal. This is not surprising since the size
of the interference term depends on our lack of knowledge of which
condensate a given detected atom comes from. If the net detection rate of
one condensate is larger than the other, then we know that the detected atom
is more likely to have originated from this particular condensate. Unlike
the previous two cases, the maximum visibility depends on $N$ the initial
total number of atoms in the condensates. The maximum visibility for the
special case of equal number and detection rate for each condensate we have 
\begin{equation}
\left[ {\cal V}_{{\rm Fock}}\right] _{\gamma _{1}=\gamma _{2},n_{1}=n_{2}}=%
\frac{1}{2\left( 1-1/N\right) }.
\end{equation}
In the limit of large $N$ the conditional visibility approaches that of the
Poissonian mixture which has the value of one half. For practical purposes
where the number of atoms in each condensate is well over one thousand, the
cases of a Poissonian mixture and of initial Fock states are
indistinguishable. We plot these conditional visibilities for the different
initial conditions as a function of the ratio of initial net detection
rates, $\Gamma \bar{n}_{1}/\bar{n}_{2}$ between the two condensates in
Fig.\thinspace (\ref{fig01}). To see how the maximum visibility changes as
we alter the width of these initial distributions, let us also consider some
arbitrary Gaussian distribution with a variance $\sigma ^{2}$ and mean $%
\overline{n}.$ The conditional visibility when both condensates start with
this Gaussian distribution and equal detection rates is 
\begin{equation}
{\cal V}_{{\rm Gaussian}}(\overline{n},\sigma ^{2})=\frac{\overline{n}^{2}}{%
\sigma ^{2}+2\overline{n}^{2}-\overline{n}}.  \label{vis_gauss}
\end{equation}
This is approximately equal to $\left( 2+\sigma ^{2}/\overline{n}^{2}\right)
^{-1}$ for large $\overline{n}.$ For wide distributions $\sigma ^{2}\gg 
\overline{n}^{2},$ we see that the conditional visibility tends to zero.
Conversely for narrow distributions where $\sigma ^{2}\ll \overline{n}^{2},$
the visibility becomes approximately one half for large $\overline{n}$. In
the special case of $\sigma ^{2}=\overline{n}$ where we approximate the
Poisson distribution by a Gaussian, Eq.\thinspace (\ref{vis_gauss}) yields a
visibility of one half which is consistent with the value obtained from the
expression for the Poissonian visibility given by Eq. (\ref{vis_poisson}).

\subsection{Conditional probability density after $m$ detections\label%
{section_many}}

The conditional probability density $p\left( x|x_{1}\right) $ displayed by
Eq.\thinspace (\ref{eqn_vis}) can be generalized to an expression governing
the probability density of $x$ given the previous $m$ measurements $\left\{
x_{1}\ldots x_{m}\right\} .$ We can write the operator representing the
cumulative effect of $m$ detections as 
\begin{eqnarray}
\hat{\Psi}\left( x_{m}\right) \hat{\Psi}\left( x_{m-1}\right) \ldots \hat{%
\Psi}\left( x_{1}\right)  &=&\prod_{j=1}^{m}\left( \hat{a}_{1}+\sqrt{\Gamma }%
\hat{a}_{2}e^{-i\phi _{j}}\right)   \label{multi_opr} \\
&=&\sum_{k=0}^{m}\pi _{k}^{\left( m\right) }\left( \phi _{1},\ldots ,\phi
_{m}\right)   \nonumber \\
&&\times \,\hat{a}_{1}^{m-k}\hat{a}_{2}^{k}\Gamma ^{k/2}
\end{eqnarray}
where we define $\phi _{k}\equiv \phi \left( x_{k}\right) $ for notational
convenience and the coefficients $\pi _{k}^{\left( m\right) }\left( \phi
_{1},\ldots ,\phi _{m}\right) $ can be found by computing the power series
expansion 
\begin{equation}
\prod_{j=1}^{m}\left( 1+ze^{-i\phi _{j}}\right) =\sum_{k=0}^{m}\pi
_{k}^{\left( m\right) }\left( \phi _{1},\ldots ,\phi _{m}\right) z^{k}.
\label{Pi_define}
\end{equation}
They satisfy the recursion relation 
\begin{equation}
\pi _{k}^{\left( m+1\right) }=\pi _{k}^{\left( m\right) }\left( 1-\delta
_{k,m+1}\right) +\pi _{k-1}^{\left( m\right) }\left( 1-\delta _{k,0}\right)
e^{-i\phi _{m+1}}
\end{equation}
where we have used the notation $\pi _{k}^{\left( m\right) }\equiv \pi
_{k}^{\left( m\right) }\left( \phi _{1},\ldots ,\phi _{m}\right) $ for
brevity. In a numerical simulation, the product (\ref{Pi_define}) can be
updated after every atomic detection by carrying out polynomial
multiplication. The un-normalized state vector after applying the above
operator to the initial state $|n_{1},n_{2}\rangle $ is 
\begin{eqnarray}
|\varphi _{m}\rangle  &=&\sum_{k=0}^{m}\sqrt{\frac{n_{1}!n_{2}!}{\left(
n_{1}-m+k\right) !\left( n_{2}-k\right) !}}  \nonumber \\
&&\times \pi _{k}^{\left( m\right) }\Gamma ^{k/2}|n_{1}-m+k,n_{2}-k\rangle .
\end{eqnarray}
Let us now consider the un-normalized wave-function after the $\left(
m+1\right) $'th detection 
\begin{equation}
|\varphi _{m+1}\left( \phi \right) \rangle =\left( \hat{a}_{1}+\sqrt{\Gamma }%
\hat{a}_{2}e^{-i\phi }\right) |\varphi _{m}\rangle 
\end{equation}
where we have explicitly shown the $\phi $ dependence. The joint probability
of $m+1$ detections at the $\left\{ x_{1},\ldots ,x_{m},x\right\} $
positions is 
\begin{eqnarray}
&&p^{\left( m+1\right) }(x_{1},\ldots ,x_{m},x)  \nonumber \\
&=&{\cal N}^{\left( m+1\right) }{\bf Tr}\left\{ \rho \hat{\psi}^{\dag
}(x_{1})\ldots \hat{\psi}^{\dag }(x_{m})\hat{\psi}^{\dag }(x)\right.  
\nonumber \\
&&\left. \times \hat{\psi}(x)\hat{\psi}(x_{m})\ldots \hat{\psi}\left(
x_{1}\right) \right\}  \\
&=&{\cal N}^{\left( m+1\right) }\sum_{n_{1}=0}^{\infty
}\sum_{n_{2}=0}^{\infty }P_{n_{1},n_{2}}\langle \varphi _{m+1}\left( \phi
\right) |\varphi _{m+1}\left( \phi \right) \rangle ,
\end{eqnarray}
where 
\begin{eqnarray}
&&\langle \varphi _{m+1}\left( \phi \right) |\varphi _{m+1}\left( \phi
\right) \rangle   \nonumber \\
&=&\frac{n_{1}!}{\left( n_{1}-m-1\right) !}+\sum_{k=0}^{m-1}\frac{\Gamma
^{k+1}n_{1}!n_{2}!}{\left( n_{1}-m+k\right) !\left( n_{2}-k-1\right) !} 
\nonumber \\
&&\times \left| \pi _{k}^{\left( m\right) }e^{-i\phi }+\pi _{k+1}^{\left(
m\right) }\right| ^{2}+\frac{\Gamma ^{m+1}n_{2}!}{\left( n_{2}-m-1\right) !}.
\end{eqnarray}
The conditional probability density is thus 
\begin{eqnarray}
&&p\left( x|x_{1},\ldots ,x_{m}\right)   \nonumber \\
&=&N^{\left( m\right) }\left[ \langle n_{1}\left( n_{1}-1\right) \ldots
\left( n_{1}-m\right) \rangle \right.  \nonumber \\
&&+\sum_{k=0}^{m-1}\langle n_{1}\left( n_{1}-1\right) \ldots \left(
n_{1}-m+k+1\right) \rangle   \nonumber \\
&&\times \langle n_{2}\left( n_{2}-1\right) \ldots \left( n_{2}-k\right)
\rangle \Gamma ^{k+1}\left| \pi _{k}^{\left( m\right) }e^{-i\phi }+\pi
_{k+1}^{\left( m\right) }\right| ^{2}  \nonumber \\
&&\left. +\Gamma ^{m+1}\langle n_{2}\left( n_{2}-1\right) \ldots \left(
n_{2}-m\right) \rangle \right]   \label{cond_prob}
\end{eqnarray}
where 
\begin{equation}
N^{\left( m\right) }=\frac{{\cal N}^{\left( m+1\right) }}{p^{\left( m\right)
}(x_{1},\ldots ,x_{m})}
\end{equation}
is a $x$ independent normalization factor. The angle brackets denotes the
sum over the probability distribution $P_{n_{1},n_{2}}$.

For different initial mixtures, either a thermal or a Poissonian
distribution, we obtain different relationships between the higher order
moments and the first (i.e. the mean). The relationship for the thermal case
is 
\begin{equation}
\langle n\left( n-1\right) \ldots \left( n-k\right) \rangle _{{\rm thermal}%
}=\left( k+1\right) !\,\overline{n}^{k+1}  \label{rel_thermal}
\end{equation}
and 
\begin{equation}
\langle n\left( n-1\right) \ldots \left( n-k\right) \rangle _{{\rm Poisson}}=%
\overline{n}^{k+1}  \label{rel_poisson}
\end{equation}
for the Poissonian distribution. Using these properties we obtain the
conditional probability density for the thermal distribution 
\begin{eqnarray}
p_{{\rm Poisson}}\left( x|x_{1},\ldots ,x_{m}\right) &=&N_{{\rm Poisson}%
}^{\left( m\right) }\left[ \bar{n}_{1}^{m+1}+\left( \Gamma \bar{n}%
_{2}\right) ^{m+1}\right.  \nonumber \\
&&+\sum_{k=0}^{m-1}{\cal A}_{k}^{\left( {\rm Poisson}\right) }  \nonumber \\
&&\left. \times \left| \pi _{k}^{\left( m\right) }e^{-i\phi }+\pi
_{k+1}^{\left( m\right) }\right| ^{2}\right]  \label{cond_poisson}
\end{eqnarray}
and 
\begin{eqnarray}
p_{{\rm thermal}}\left( x|x_{1},\ldots ,x_{m}\right) &=&N_{{\rm thermal}%
}^{\left( m\right) }\left\{ \left( m+1\right) !\left[ \bar{n}%
_{1}^{m+1}\right. \right.  \nonumber \\
&&\left. +\left( \Gamma \bar{n}_{2}\right) ^{m+1}\right] +\sum_{k=0}^{m-1}%
{\cal A}_{k}^{\left( {\rm thermal}\right) }  \nonumber \\
&&\left. \times \left| \pi _{k}^{\left( m\right) }e^{-i\phi }+\pi
_{k+1}^{\left( m\right) }\right| ^{2}\right\}  \label{cond_thermal}
\end{eqnarray}
for the Poissonian distribution. Note that the $x$ independent normalization
factors are now calculated with respect to the relevant density operator,
i.e. Poisson for $N_{{\rm Poisson}}^{\left( m\right) }$ and thermal for $N_{%
{\rm thermal}}^{\left( m\right) }.$ Both the thermal and Poisson
distribution's conditional probability densities have a similar form with
different weighting factors ${\cal A}_{k}$ given by 
\begin{equation}
{\cal A}_{k}^{\left( {\rm Poisson}\right) }=\bar{n}_{1}^{m-k}\bar{n}%
_{2}^{k+1}\Gamma ^{k+1}
\end{equation}
and 
\begin{equation}
{\cal A}_{k}^{\left( {\rm thermal}\right) }=\left( m-k\right) !\left(
k+1\right) !{\cal A}_{k}^{\left( {\rm Poisson}\right) }.
\end{equation}
The generalized conditional probability distributions Eq. (\ref{cond_poisson}%
) and Eq. (\ref{cond_thermal}) groups all the measurement dependent terms,
the $\left\{ \phi _{1},\ldots ,\phi _{m}\right\} $ detections, within the $%
\left| \ldots \right| $ brackets. Numerical calculations are readily
obtainable since we need only to simulate the expression within the $\left|
\ldots \right| $ brackets. We shall come back to this point and explain in
more detail the numerical simulation process when we describe the numerical
results in section \ref{Results}.

\subsection{Analytical results for a large number of detections\label%
{section 2}\label{analytic}}

\subsubsection{Poissonian mixtures}

In the case of Poissonian mixtures the expression (\ref{cond_poisson}) can
be made more explicit and analytical results can be extracted. As shown by
Cirac {\it et. al.} \cite{Cir} it is useful to represent the state after $m$
detections in the form of a P-representation\thinspace 
\begin{eqnarray}
\rho _{m} &=&\int \frac{d\psi }{2\pi }f_{m}\left( \psi \right) \int \frac{%
d\varphi _{1}}{2\pi }\left| \alpha _{1}e^{i\varphi _{1}}\right\rangle
\left\langle \alpha _{1}e^{i\varphi _{1}}\right|  \nonumber \\
&&\otimes \left| \alpha _{2}e^{i\left( \varphi _{1}+\psi \right)
}\right\rangle \left\langle \alpha _{2}e^{i\left( \varphi _{1}+\psi \right)
}\right|  \label{density}
\end{eqnarray}
with 
\begin{equation}
\alpha _{1}=\sqrt{\bar{n}_{1}},\alpha _{2}=\sqrt{\bar{n}_{2}}.
\end{equation}
In writing Eq. (\ref{density}) we are assuming that the observation time $%
t_{m}$ satisfies $\gamma _{i}t_{m}\ll 1$, so that only a negligible fraction
of the atoms in the condensates are counted. The initial Poissonian mixture $%
\rho _{0}$ is also of the form of Eq. (\ref{density}) with 
\begin{equation}
f_{0}\left( \psi \right) =1.  \label{f_fun01}
\end{equation}
The effect of an additional counting event on (\ref{density}) is to change 
\[
f_{m}\left( \psi \right) \rightarrow f_{m+1}\left( \psi \right) 
\]
with 
\begin{equation}
f_{m+1}\left( \psi \right) =\frac{1}{N_{m+1}}\left[ 1+\lambda \cos \left(
\phi _{m+1}-\psi \right) \right] f_{m}\left( \psi \right)  \label{f_fun02}
\end{equation}
where 
\begin{equation}
\lambda =\frac{2\sqrt{\Gamma \bar{n}_{1}\bar{n}_{2}}}{\bar{n}_{1}+\Gamma 
\bar{n}_{2}}  \label{Rvisibility}
\end{equation}
and $N_{m+1}$ is determined by normalizing $\int \frac{d\psi }{2\pi }%
f_{m+1}\left( \psi \right) =1$ after each counting event. We note that $%
0<\lambda \leq 1.$ In the following we shall assume that $\lambda <1,$ as
the case $\lambda =1$ requires a separate mathematical treatment. The
physical meaning of $f_{m}\left( \psi \right) $ is clear from Eq. (\ref
{density}): it is the probability distribution of the relative phase between
the two condensate modes. The explicit form of $f_{m}\left( \psi \right) $
as a function of $\psi $ is easily obtained from Eq. (\ref{f_fun01}) and Eq.
(\ref{f_fun02}) as\footnote{%
This formula permits us to appreciate why the case $\lambda =1$ is very
special: For $\lambda <1$ the distribution (\ref{fm}) is a positive function 
$f_{m}\left( \psi \right) >0$ everywhere. For $\lambda =1$ it has $m$ $2$%
-fold degenerate zeros, i.e. $f_{m}\left( \psi \right) $ cannot approach a
smooth function for $m\rightarrow \infty .$} 
\begin{equation}
f_{m}\left( \psi \right) =\prod_{k=1}^{m}\left[ 1+\lambda \cos \left( \phi
_{k}-\psi \right) \right] \frac{1}{N_{k}}.  \label{fm}
\end{equation}
The normalized joint probability distribution to observe the phases $\phi
_{1},\ldots ,\phi _{m+1}$ is equal to 
\begin{equation}
\int \frac{d\psi }{2\pi }\prod_{k=1}^{m+1}\left[ 1+\lambda \cos \left( \phi
_{k}-\psi \right) \right] =\prod_{k=1}^{m+1}N_{k}.
\end{equation}
Therefore the conditional probability (\ref{cond_poisson}) determined by the
ratio of $2$ joint probabilities is equal to $N_{m+1}$ and given by 
\begin{equation}
p_{{\rm Poisson}}\left( x\mid x_{1},\ldots ,x_{m}\right) =\int \frac{d\psi }{%
2\pi }\left[ 1+\lambda \cos \left( \phi -\psi \right) \right] f_{m}\left(
\psi \right) .  \label{Rpossion}
\end{equation}
where $\phi =\phi \left( x\right) $ and $f_{m}\left( \psi \right) $ depends
on the previous $m$ detections. Now we investigate the behavior of $%
f_{m}\left( \psi \right) $ as a function of $\psi $ for large $m.$ We shall
assume that it becomes a narrow distribution centered around some $m$%
-dependent maximum $\psi _{m}$ with a variance $\sigma _{m}^{2}$, i.e. we
put 
\begin{equation}
f_{m}\left( \psi \right) =\frac{1}{\sqrt{2\pi }\sigma _{m}}\exp \left[ -%
\frac{\left( \psi -\psi _{m}\right) ^{2}}{2\sigma _{m}^{2}}\right] .
\label{gaussian}
\end{equation}
We assume here that $\sigma _{m}^{2}\ll 1$, so that the $2\pi $-periodicity
of $f_{m}\left( \psi \right) $ is not in noticeable conflict with (\ref
{gaussian}).

We shall now determine the time evolution of $\psi _{m}$ and $\sigma
_{m}^{2} $, using Eq. (\ref{f_fun02}), and show that our assumptions are
self-consistent in that indeed $\sigma _{m}^{2}\ll 1$ for $m\gg 1.$ From Eq.
(\ref{fm}) we obtain by taking the logarithm 
\begin{equation}
\ln f_{m}\left( \psi \right) =\sum_{k=1}^{m}\ln \left[ 1+\lambda \cos \left(
\phi _{k}-\psi \right) \right] +c
\end{equation}
where the constant $c$ depends on the $\phi _{m}$ but not on $\psi .$ The
maximum $\psi _{m}$ of $\ln f_{m}\left( \psi \right) $ must therefore
satisfy 
\begin{equation}
\sum_{k=1}^{m}\frac{\lambda \sin \left( \phi _{k}-\psi _{m}\right) }{%
1+\lambda \cos \left( \phi _{k}-\psi _{m}\right) }=0.
\end{equation}
The evolution of the maximum is obtained similarly by taking the logarithm
of Eq. (\ref{f_fun02}) and using Eq. (\ref{gaussian}). We find 
\begin{equation}
-\frac{\left( \psi _{m+1}-\psi _{m}\right) }{\sigma _{m+1}^{2}}+\frac{%
\lambda \sin \left( \phi _{m+1}-\psi _{m}\right) }{1+\lambda \cos \left(
\phi _{m+1}-\psi _{m}\right) }=0.
\end{equation}
It is clear that $\left( \psi _{m+1}-\psi _{m}\right) $ depends on the
outcome of the $\left( m+1\right) $'th measurement. If $f_{m}\left( \psi
\right) $ is, in fact, a narrow distribution on the scale $2\pi ,$ as we
have assumed, then the probability distribution to find $\phi _{m+1}$ in
that measurement can be estimated as 
\begin{equation}
P_{m+1}\left( \phi _{m+1}\right) \simeq 1+\lambda e^{-\frac{1}{2}\sigma
_{m}^{2}}\cos \left( \phi _{m+1}-\psi _{m}\right)
\end{equation}
as follows from Eq. (\ref{Rpossion}) and Eq. (\ref{gaussian}). Based on this
we find 
\begin{eqnarray}
\left\langle \frac{\psi _{m+1}-\psi _{m}}{\sigma _{m+1}^{2}}\right\rangle
&=&0,  \nonumber \\
\left\langle \frac{\left( \psi _{m+1}-\psi _{m}\right) ^{2}}{\sigma
_{m+1}^{4}}\right\rangle &=&\lambda ^{2}\int \frac{d\phi }{2\pi }\frac{\sin
^{2}\phi }{1+\lambda \cos \phi }  \nonumber \\
&&-\frac{\lambda ^{3}}{2}\sigma _{m}^{2}\int \frac{d\phi }{2\pi }\frac{\cos
\phi \sin ^{2}\phi }{\left( 1+\lambda \cos \phi \right) ^{2}}  \nonumber \\
&=&1-\sqrt{1-\lambda ^{2}}  \nonumber \\
&&+\frac{1}{2}\sigma _{m}^{2}\frac{\left( 1-\sqrt{1-\lambda ^{2}}\right) ^{2}%
}{\sqrt{1-\lambda ^{2}}}
\end{eqnarray}
where we used $\sigma _{m}^{2}\ll 1$ to expand to first order $\exp \left(
-\sigma _{m}^{2}/2\right) =1-\sigma _{m}^{2}/2.$ We conclude that, if $%
\lambda $ is bounded away from $1$, for small variance of the phase
distribution the variance in the jitter of the average phase $\psi _{m}$ in
subsequent measurements is of the order of the {\em square} of the variance $%
\sigma _{m}^{2}$ in the phase distribution, i.e. the position of the maximum
is very stable and may be considered as fixed in the limit we consider.

Turning to the time evolution of the variance we obtain again using Eq. (\ref
{gaussian}) on the right hand side of Eq. (\ref{f_fun02}), taking the
logarithm and the second derivative with respect to $\psi $%
\begin{equation}
\frac{1}{\sigma _{m+1}^{2}}=\frac{1}{\sigma _{m}^{2}}+\frac{\lambda
^{2}+\lambda \cos \left( \phi _{m+1}-\psi _{m}\right) }{\left[ 1+\lambda
\cos \left( \phi _{m+1}-\psi _{m}\right) \right] ^{2}}.
\end{equation}
It shows that typically the inverse variance grows according to 
\begin{equation}
\frac{1}{\sigma _{m+1}^{2}}-\frac{1}{\sigma _{m}^{2}}=O\left( 1\right)
\end{equation}
which is the reason why the variance itself indeed becomes small. Averaging
as before we obtain for $\lambda $ bounded away from $1$%
\begin{equation}
\left\langle \frac{1}{\sigma _{m}^{2}}\right\rangle =\left[ \left( 1-\sqrt{%
1-\lambda ^{2}}\right) +O\left( \sigma _{m}^{2}\right) \right] m+{\rm const.}
\label{OneOver_m}
\end{equation}
which shows more explicitly how the inverse variance grows on the average
and becomes large for large $m$.

Finally, we relate the phase distribution $f_{m}\left( \psi \right) $ after $%
m$ measurements to the observed interference pattern. Experimentally, a
phase distribution may be extracted by fitting the observed interference
pattern, normalized with respect to its constant part, to the expected
density Eq. (\ref{Rpossion}) after $m$ counts 
\begin{equation}
\int \frac{d\psi }{2\pi }\left[ 1+\lambda \cos \left( \phi -\psi \right)
\right] f_{m}\left( \psi \right) =1+\lambda ^{\prime }\cos \left( \phi -\psi
_{m}\right)  \label{xpt1}
\end{equation}
with 
\begin{equation}
\lambda ^{\prime }=\lambda e^{-\frac{1}{2}\sigma _{m}^{2}}.  \label{xpt2}
\end{equation}
Here we used the Gaussian form of $f_{m}\left( \psi \right) $ assumed in Eq.
(\ref{gaussian}). Equations (\ref{xpt1}), (\ref{xpt2}) allow us to extract
numbers for $\psi _{m}$ and $\sigma _{m}^{2}$. A weak link in the argument
leading to Eq. (\ref{xpt1}) might seem to be the fact that the conditional
probability (\ref{Rpossion}) is proportional to the expectation value of the
density in the condensates, while what we really need is a normalized
measure of the density of the counted atoms. However, all that is really
required is that these two quantities should be proportional to each other,
which is an assumption implicitly already made when assuming that the
counting rate is proportional to the number operator in the condensate.

\subsubsection{Thermal mixtures}

If the initial states of the two condensates are thermal mixtures we may
adapt the results for the Poissonian mixtures as follows. The thermal
initial state can be represented as a mixture of Poissonian states via 
\begin{eqnarray}
\rho _{0} &=&\frac{1}{Z_{1}Z_{2}}\sum_{n_{1},n_{2}}e^{-\beta \left(
n_{1}+n_{2}\right) }\left| n_{1}\right\rangle \left\langle n_{1}\right|
\otimes \left| n_{2}\right\rangle \left\langle n_{2}\right|  \nonumber \\
&=&\frac{1}{{\cal N}}\int d^{2}\alpha _{1}\int d^{2}\alpha _{2}\exp \left( -%
\frac{\left| \alpha _{1}\right| ^{2}}{\bar{n}_{1}}-\frac{\left| \alpha
_{2}\right| ^{2}}{\bar{n}_{2}}\right) \left| \alpha _{1}\right\rangle
\left\langle \alpha _{1}\right|  \nonumber \\
&&\otimes \left| \alpha _{2}\right\rangle \left\langle \alpha _{2}\right| 
\nonumber \\
&=&\frac{1}{{\cal N}^{\prime }}\int_{0}^{\infty }dx_{1}\int_{0}^{\infty
}dx_{2}\exp \left( -\frac{x_{1}}{\bar{n}_{1}}-\frac{x_{2}}{\bar{n}_{2}}%
\right)  \nonumber \\
&&\times \int \frac{d\psi }{2\pi }f_{0}\left( \psi \right) \int \frac{%
d\varphi _{1}}{2\pi }\left| \sqrt{x_{1}}e^{i\varphi _{1}}\right\rangle
\left\langle \sqrt{x_{1}}e^{i\varphi _{1}}\right|  \nonumber \\
&&\otimes \left| \sqrt{x_{2}}e^{i\left( \varphi _{1}+\psi \right)
}\right\rangle \left\langle \sqrt{x_{2}}e^{i\left( \varphi _{1}+\psi \right)
}\right|  \label{thermal_den}
\end{eqnarray}
with the inverse temperature proportional to $\beta $ and normalization
factors ${\cal N}$ and ${\cal N}^{\prime }$. The evolution of this initial
state due to the counting of atoms can now be obtained by using the
evolution for Poissonian mixtures for fixed Poissonian averages $x_{1},x_{2}$
under the integrals over $x_{1},x_{2}$. We obtain with $f_{0}\left( \psi
\right) =1,$ using Eq. (\ref{fm}) for 
\begin{equation}
\lambda =\lambda \left( x_{1},x_{2}\right) =\frac{2\sqrt{\Gamma x_{1}x_{2}}}{%
x_{1}+\Gamma x_{2}}  \label{lambda}
\end{equation}
\begin{equation}
f_{m}\left( \psi \right) \sim \prod_{k=1}^{m}\left[ x_{1}+\Gamma x_{2}+2%
\sqrt{\Gamma x_{1}x_{2}}\cos \left( \phi _{k}-\psi \right) \right] .
\end{equation}
The optimal visibility $\lambda $ thus varies according to Eq. (\ref{lambda}%
) for different members of the ensemble whose $x_{1},x_{2}$ values are
exponentially distributed according to Eq. (\ref{thermal_den}). The average
visibility becomes 
\begin{eqnarray*}
\bar{\lambda} &=&\frac{2\sqrt{\Gamma }\left\langle \sqrt{x_{1}x_{2}}%
\right\rangle }{\left\langle x_{1}\right\rangle +\Gamma \left\langle
x_{2}\right\rangle } \\
&=&\frac{\pi }{4}\frac{2\sqrt{\Gamma \left\langle x_{1}\right\rangle
\left\langle x_{2}\right\rangle }}{\left\langle x_{1}\right\rangle +\Gamma
\left\langle x_{2}\right\rangle }
\end{eqnarray*}
which may be written as\footnote{%
This is consistent with the visibility defined by Eq. (\ref{Rvisibility})
previously since $\lambda \left( \left\langle x_{1}\right\rangle
,\left\langle x_{2}\right\rangle \right) \equiv \lambda .$} 
\begin{equation}
\bar{\lambda}=\frac{\pi }{4}\lambda .
\end{equation}
For equal average counting rates 
\[
\bar{n}_{1}=\Gamma \bar{n}_{2} 
\]
the average visibility for initially thermal mixtures is simply 
\begin{equation}
\bar{\lambda}=\frac{\pi }{4},  \label{Pi_over4}
\end{equation}
which agrees well with the result of the numerical simulations to be
presented below.

\subsection{Numerical results\label{Results}}

The form of the expressions for the conditional probabilities, Eq. (\ref
{cond_poisson}) and Eq. (\ref{cond_thermal}) of section (\ref{section_many})
can be readily applied to stochastic simulations of atom detection which
generates interference patterns. The procedure follows the spirit of
Javanainen and Yoo's work; a random number $\phi _{1}$ is generated for the
initial atom detection (since we know the initial conditional probability
distribution is uniform), then it is used to calculate the conditional
probability density $p\left( \phi |\phi _{1}\right) $ so that the second
detection $\phi _{2}$ can be generated by randomly selecting a value
according to $p\left( \phi |\phi _{1}\right) .$ This process is repeated to
generate $m$ atom detections which are binned and displayed as an
interference pattern.

Examples of these stochastically generated interference patterns are
displayed in fig.\thinspace \ref{fig02} as histograms of the raw output from
the simulations, the sequence of detected positions $\left\{ \phi
_{1},\ldots ,\phi _{m}\right\} $ are sorted into $25$ bins plotted as
circles. Figure \thinspace \ref{fig02}(a) shows the interference pattern for
an initial thermal mixture which has a visibility of $0.79$ whereas in
fig.\thinspace \ref{fig02}(b) we have a visibility of $0.97$ for the
Poissonian mixture. These visibilities are calculated via a least-squares
fit of the form $1+\beta \cos \left( 2\pi +\phi \right) $ shown as the solid
curves in fig. \ref{fig02}(a) and (b). In both cases we have simulated $500$
detections.

The numerical simulations used to generate the histograms in fig.\thinspace 
\ref{fig02}(a) and (b) also calculate the conditional probability
distributions before each detection as these are necessary to calculate the
location of the detected atom. The evolution of these conditional
distributions gives an insight into the build up of the interference pattern 
\cite{Wong}. The visibility of the complete interference pattern (as
calculated from the least-square fit) can be considered as an average over
these conditional visibilities. Graphs of the conditional visibility as a
function of atom detections typically approach a value close to $1$ within $%
100$ detections and stay at that value thereafter. Thus the conditional
visibility after $m$ detections can be thought of as a good approximation of
the visibility of the complete interference pattern for values of $m>100.$
We have not displayed these graphs of stochastically generated sequences
since they possess fluctuations about a generic shape which is a property of
all such sequences. We plot the average conditional visibility over many
such sequences in fig. \ref{fig03} and fig. \thinspace \ref{fig04} for
Poisson and thermal mixtures respectively. The ratio $\Gamma \bar{n}_{2}/%
\bar{n}_{1}$ was set at $1$ with the average performed over $1000$ runs (we
shall hereby refer to an individual sequence of detections as a single
``run'' for convenience). The shaded regions around the average conditional
visibility displayed as a solid line depict the extent of the fluctuations
for an individual run. The boundaries of this shaded region corresponds to
the upper and lower quartiles respectively, $25\%$ of the data lies below
the lower quartile whereas $25\%$ lies above the upper quartile, thus the
probability that the fluctuations lie within this region is $50\%.$ These
fluctuations are much larger for the thermal mixtures in comparison with
those of the Poissonian mixtures.

The maximum value of the visibility in fig. \ref{fig03} occurs at $500$
atoms detected with a value of $0.999$ ($3$ sig. fig.). The value of the
visibility obtained from Eq. (\ref{xpt2}) using $\sigma _{m}^{2}\sim 1/m$ is 
$0.999$ ($3$ sig. fig.) which agrees very well with the simulation. In the
case of the thermal mixture, fig. \ref{fig04}, the maximum value of the
visibility is $0.777$ again at $500$ detections. This is close to the
analytical value of $\pi /4\approx 0.785$ predicted by Eq. (\ref{Pi_over4}).

The large difference in the fluctuations between the two cases is due to
their differing degree of sensitivity to particular runs. In the Poisson
case, it is relatively insensitive since the majority of the terms in the $%
\left| \ldots \right| $ brackets of Eq. (\ref{cond_poisson}) depends on the $%
\left\{ \phi _{1},\ldots ,\phi _{m}\right\} $ detections fairly equally
whereas in the thermal case we see there is an additional factorial factor
in front of the $\left| \ldots \right| $ brackets which favors $k$ values at
the ends (close to zero and $m-1$). This enhances the sensitivity of the
visibility upon particular combinations of $\phi _{k}$ thus the sensitivity
is high. Alternatively, the large fluctuations of the visibility in the
thermal case is not surprising when we consider the analytic treatment of
section (\ref{analytic}{\em 2}). The thermal state was represented as a
mixture of Poissonian states so that the averaged conditional visibility has
an additional average over the mixture of Poissonian states.

The variance for the Poisson case can be estimated from fig. \ref{fig03} by
using Eq. (\ref{xpt2}) since the fluctuations are small. When the variance
are small compared to $1$, the variance is 
\begin{equation}
\sigma _{m}^{2}=2\left( \lambda -\lambda ^{\prime }\right)
\end{equation}
where $\lambda ^{\prime }$ is the numerical averaged visibility and in the
case of equal counting rates $\lambda =1$. The variance for the Poisson case
is graphed as the solid curve in fig. \ref{fig05} with the dashed curve
displaying the $\sigma _{m}^{2}\sim 1/m$ relationship predicted by Eq. (\ref
{OneOver_m}). Note that we have not plotted variances below $50$ detections
since Eq. (\ref{xpt2}) is only valid for number of detections $m\gg 1.$ As
expected, the agreement between the two curves becomes better as more atoms
are detected. We cannot obtain a good estimate of the variance in the same
manner for the thermal case because the fluctuations in the visibility are
large.

We will show in the appendix that the results for measurements induced phase
distribution for initial number states are the same as those obtained for
the Poissonian mixtures. Numerical simulations of the time evolution and
effect of detections on the wave-function of the condensates have been
performed\cite{Wong} where the variance can be calculated directly from this
wave-function. This is possible in the simpler case of initial number
states. Because the phase is periodic it is convenient to use the following
measure of the spread of the phase distribution: 
\begin{equation}
\delta \phi =1-\left\langle \text{{\bf cos}}\Delta \phi \right\rangle
^{2}-\left\langle \text{{\bf sin}}\Delta \phi \right\rangle ^{2}
\label{measure}
\end{equation}
where the trigonometric operators are defined as 
\begin{equation}
\text{{\bf cos}}\Delta \phi =\frac{1}{2}\left( \widehat{e}^{i\Delta \phi }+%
\widehat{e}^{-i\Delta \phi }\right)  \label{cos}
\end{equation}
and 
\begin{equation}
\text{{\bf sin}}\Delta \phi =\frac{1}{2\imath }\left( \widehat{e}^{i\Delta
\phi }-\widehat{e}^{-i\Delta \phi }\right) .  \label{sin}
\end{equation}
with the SG phase operators $\widehat{e}^{i\Delta \phi }$ defined for the
relative phase between the two condensates. This measure ranges from zero to
one with values close to zero agreeing well with the actual variance. Since
our expressions are only valid for small variances, this measure is very
useful as an estimate of the variance. Figure \ref{fig06} plots three curves
of $\delta \phi ,$ each one with differing relative count rates between the
condensates ($\Gamma N_{2}/N_{1}$) where $N_{1}$ and $N_{2}$ are the initial
atom number in the first and second condensate respectively. Analytical
predictions of the gradient for each relative count rate are obtained from
Eq. (\ref{OneOver_m}). The solid, dashed and dash-dotted curves of fig. \ref
{fig06} corresponds to $\Gamma N_{2}/N_{1}$ ratios of $1,$ $1/2,$ and $1/4$
which are predicted to have gradients $1,$ $2/3,$ and $2/5$ respectively.
The numerical gradient was calculated for points from $20$ detections
onwards, the points below $20$ were ignored because the predictions required 
$m\gg 1$. The curves clearly show larger curvature in this region in
comparison to points after $20.$ We obtained numerical gradients of $0.97,$ $%
0.66$, and $0.41$ for the $\Gamma N_{2}/N_{1}$ ratios; $1,$ $1/2,$ and $1/4$%
. This good agreement between the analytic and numerical gradients verifies
the relationship of the visibility $\lambda $ with the relative counting
rates, Eq. (\ref{Rvisibility},\ref{lambda},\ref{append_lambda}). Note that
if we had graphed $\delta \phi $ itself instead of its inverse, we would had
seen a curve shaped like that of fig. \ref{fig05}, hence the straightness of
the curves of fig. \ref{fig06} indicates the accuracy of the $1/m$
prediction of the previous section (\ref{analytic})\footnote{%
Note that section (\ref{analytic}) consider Poisson and thermal states but
fig. \ref{fig06} show results from simulation of initial number states,
however we will show in the appendix that the width of the phase
distribution has the same relationship with the number of detections $m$ as
the Poisson state in the limit $1\ll m\ll N_{1},N_{2}.$}.

\section{Summary}

We have analyzed in detail the build up in quantum coherence between two
Bose-Einstein condensates which are initially in a thermal or Poisson state.
Interference patterns are produced via spatial atom detections which
establishes an arbitrary but fixed relative phase between the condensates.
In the regime where the total number of atoms detected is only a negligible
fraction of the atoms in the condensates although the actual number of
detections is much greater than $1$, we find that the visibility of the
interference pattern for the Poisson distribution depends on the relative
counting rates of each condensate with a maximum of one for equal rates. In
the thermal case, the visibility becomes a stochastic variable which varies
from run to run around an averaged value determined again by the relative
counting rates but the average has a maximum of $\pi /4$ for equal rates in
good agreement with our numerical simulations. The inverse variance of the
phase distribution grows linearly with the number of detected atoms $m$ in
the regime where $1\ll m\ll N_{1},N_{2}.$ This has been shown analytically
for the Poisson state and this relationship also holds true for the thermal
state since we may write the thermal state as a mixture of Poisson states.
In the appendix we have shown that the results for the initial number states
follows those derived for the Poisson state. In particular, the inverse
variance is proportional to $m,$ which has been numerically verified.

This research was supported by the University of Auckland Research
Committee, the New Zealand Lottery Grants Board and the Marsden Fund of the
Royal Society of New Zealand.

\appendix

\section{Initial number states}

Let us consider the case of initial number states and compare the results
with those of initial Poissonian mixtures. The initial number state 
\begin{equation}
\left| \psi _{0}\right\rangle =\left| N_{1}\right\rangle \left|
N_{2}\right\rangle
\end{equation}
evolves into 
\begin{equation}
\left| \psi _{m}\right\rangle =\sum_{n=0}^{m}C_{n}\left( m\right) \left|
N_{1}-n\right\rangle \left| N_{2}-m+n\right\rangle
\end{equation}
where $C_{n}\left( 0\right) =\delta _{n,0}$. We shall assume $\gamma
_{1}N_{1}>\gamma _{2}N_{2}$ and that the parameter 
\begin{equation}
\lambda _{N}=\frac{2\sqrt{\Gamma N_{1}N_{2}}}{N_{1}+\Gamma N_{2}}
\label{append_lambda}
\end{equation}
is bounded away from $1$. We shall only consider the case 
\begin{equation}
1\ll m\ll N_{1},N_{2}.
\end{equation}
Then the evolution of the $C_{n}\left( m\right) $ under the atom counting
process is 
\begin{eqnarray}
C_{n}\left( m+1\right) &\simeq &\sqrt{N_{1}}\left( 1-\delta _{n,0}\right)
C_{n-1}\left( m\right)  \nonumber \\
&&+\sqrt{\Gamma N_{2}}\left( 1-\delta _{n,m+1}\right) e^{-i\phi
_{m+1}}C_{n}\left( m\right) .  \label{recursion}
\end{eqnarray}
We introduce a phase-representation by the Fourier transform 
\begin{eqnarray}
F_{m}\left( \varphi \right) &=&\sum_{n=0}^{m}C_{n}\left( m\right)
e^{-in\varphi }  \nonumber \\
C_{n}\left( m\right) &=&\int \frac{d\varphi }{2\pi }e^{in\varphi
}F_{m}\left( \varphi \right) .  \label{FT}
\end{eqnarray}
The number density after $m$ measurements is then 
\begin{eqnarray*}
P_{m}\left( x\right) &\simeq &\left\langle \psi _{m}\right| \hat{\psi}^{\dag
}\left( x\right) \hat{\psi}\left( x\right) \left| \psi _{m}\right\rangle \\
&\simeq &1+\lambda _{N}%
\mathop{\rm Re}%
\left\{ \sum_{n=0}^{m}\left( 1-\delta _{n,0}\right) \right. \\
&&\times \int \frac{d\varphi }{2\pi }\int \frac{d\varphi ^{\prime }}{2\pi }%
F_{m}^{*}\left( \varphi \right) F_{m}\left( \varphi ^{\prime }\right)
e^{-in\left( \varphi -\varphi ^{\prime }\right) } \\
&&\left. \times e^{i\left[ \varphi -\phi \left( x\right) \right] }\right\}
\end{eqnarray*}
which for $m\gg 1$ can be approximated by 
\begin{equation}
P_{m}\left( x\right) \simeq 1+\lambda _{N}\int \frac{d\varphi }{2\pi }\left|
F_{m}\left( \varphi \right) \right| ^{2}\cos \left[ \varphi -\phi \left(
x\right) \right] .
\end{equation}
This formula should be compared with Eq. (\ref{Rpossion}) for Poissonian
mixtures. It can then be seen that $\left| F_{m}\left( \varphi \right)
\right| ^{2}$ and $f_{m}\left( \phi \right) $ play identical roles. The
evolution of $F_{m}\left( \varphi \right) $ under the atom counting process
follows from Eq. (\ref{recursion}) and (\ref{FT}) 
\begin{equation}
F_{m+1}\left( \varphi \right) \simeq \left( \sqrt{N_{1}}e^{-i\varphi }+\sqrt{%
\Gamma N_{2}}e^{-i\phi _{m+1}}\right) F_{m}\left( \varphi \right)
\end{equation}
and hence, providing a normalization factor $N_{m+1}$ to keep $\left|
F_{m+1}\left( \varphi \right) \right| ^{2}$ normalized if $\left|
F_{m+1}\left( \varphi \right) \right| $ is 
\begin{equation}
\left| F_{m+1}\left( \varphi \right) \right| ^{2}\simeq \left[ 1+\lambda
_{N}\cos \left( \phi _{m+1}-\varphi \right) \right] \left| F_{m}\left(
\varphi \right) \right| ^{2}\frac{1}{N_{m+1}}
\end{equation}
which, to the accuracy we have considered here, is the same as Eq. (\ref
{f_fun02}) for Poissonian mixtures. Hence, the results for the measurement
induced phase distribution for number states in the limit $1\ll m\ll
N_{1},N_{2}$ are the same as the results for Poissonian mixtures obtained in
section \ref{section 2}.

\begin{figure}[tbp]
\caption{The conditional visibility curves for an initial number state ($%
n=20 $), Poisson and thermal states plotted as dashed, solid and dash-dotted
curves respectively. The visibility is dependent upon the ratio of the mean
counting rates between the condensates ($\Gamma {\overline{n}}_{2}/ 
\overline{n}_{1}$).}
\label{fig01}
\end{figure}

\begin{figure}[tbp]
\caption{Histogram of $500$ numerically generated atomic detections plotted
as circles. The solid curve is a least-squares fit of the form $1+\beta \cos
(2\pi x+\phi )$. A thermal initial state is shown in (a) and an initial
Poisson state is shown in (b).}
\label{fig02}
\end{figure}

\begin{figure}[tbp]
\caption{Plot of the conditional visibility averaged over one thousand runs
versus the number of atomic detections for an initial Poisson state where $%
\overline{n}_{1}=\Gamma\overline{n}_{2}$. The shaded region corresponds to
the interquartile range of the individual runs with the mean over all runs
depicted by the solid line. Therefore $50\%$ of the run lies within this
shaded region.}
\label{fig03}
\end{figure}

\begin{figure}[tbp]
\caption{Plot of the conditional visibility averaged over one thousand runs
versus the number of atomic detections for an initial thermal state where $%
\overline{n}_{1}=\Gamma\overline{n}_{2}$. The shaded region corresponds to
the interquartile range of the individual runs with the mean over all runs
depicted by the solid line. Therefore $50\%$ of the run lies within this
shaded region. We also plot a dashed line at $\pi/4$ corresponding to the
value of the average visibility predicted from the analytical work.}
\label{fig04}
\end{figure}

\begin{figure}[tbp]
\caption{Plot of variance versus the number of detected atoms for the
Poisson state shown as the solid line. The dashed line is a plot of $%
\sigma_{m}^{2}\sim 1/m$ for comparison.}
\label{fig05}
\end{figure}

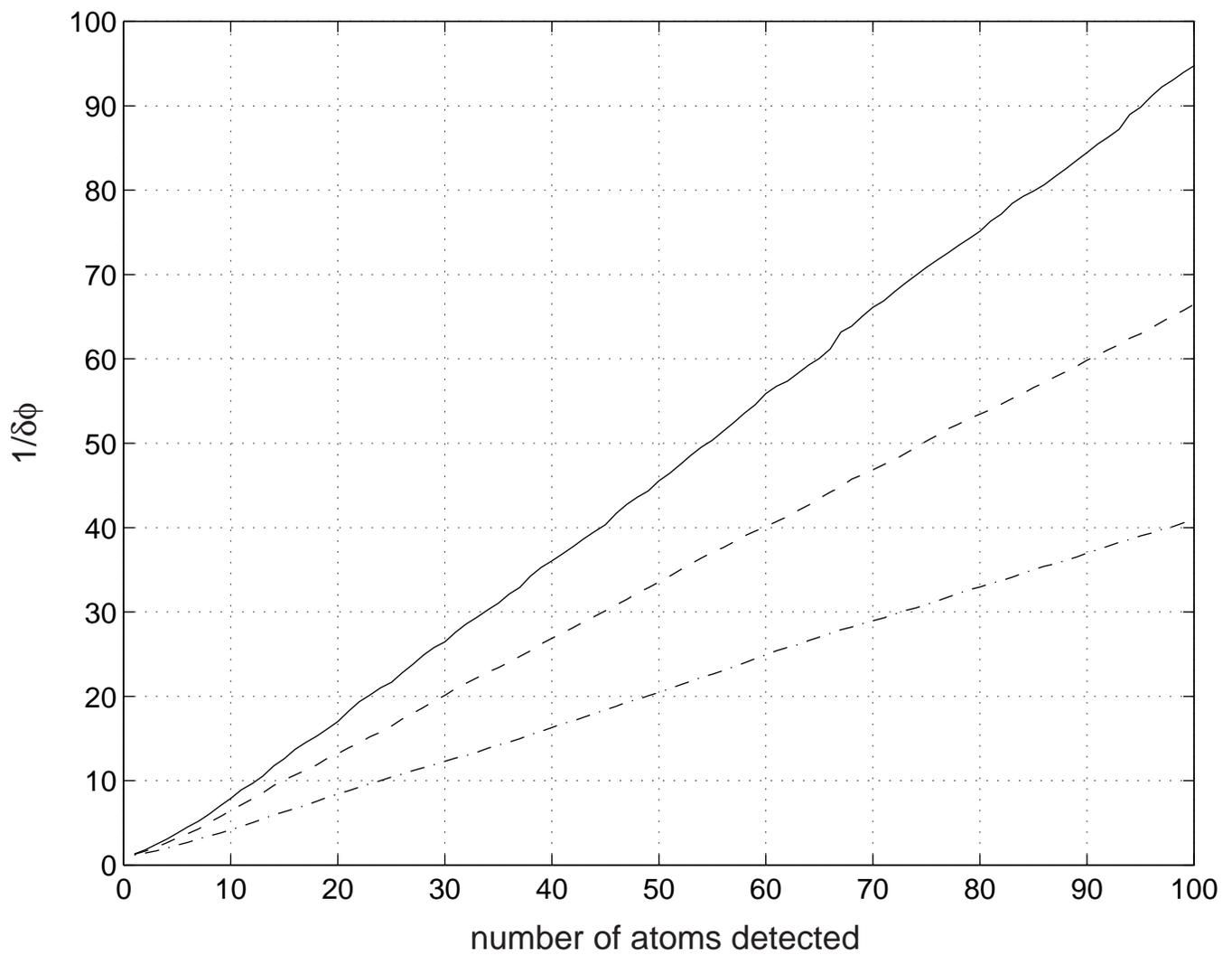
\begin{figure}[tbp]
\caption{Plot of $\delta\phi$ versus number of atomic detections with $\Gamma%
{\overline n}_{2}/\overline{n}_{1}=1, 1/2$ and $1/4$ for the solid, dashed
and dash-dotted curves respectively.}
\label{fig06}
\end{figure}

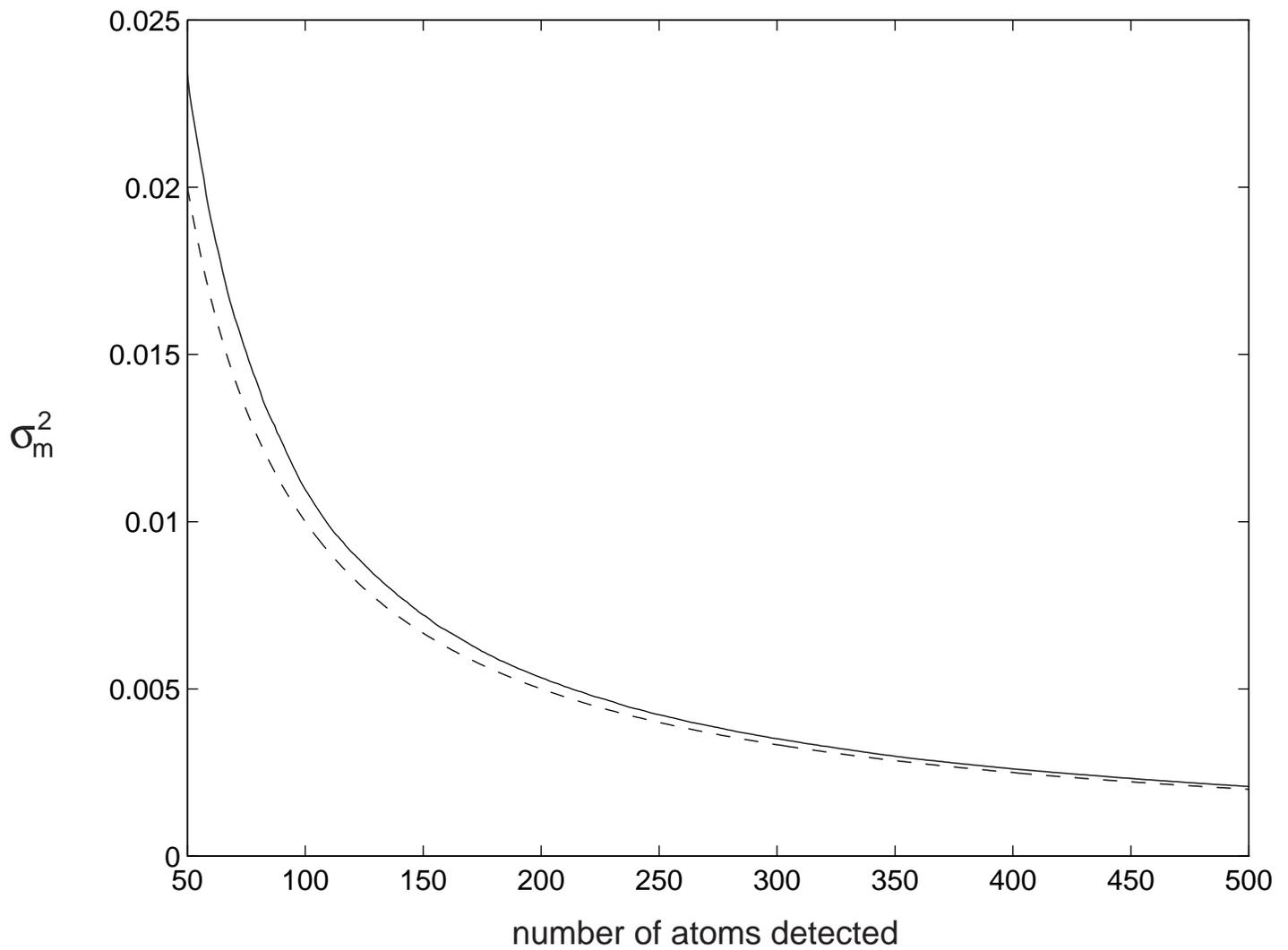
\end{document}